# Reducing Frequency Bias of Fourier Neural Operators in 3D Seismic Wavefield Simulations Through Multi-Stage Training

Qingkai Kong[1*], Caifeng Zou[2], Youngsoo Choi[1], Eric M. Matzel[1], Kamyar Azizzadenesheli[3], Zachary E. Ross[2], Arthur J. Rodgers[1], Robert W. Clayton[2]

[1]Lawrence Livermore National Laboratory, Livermore, CA, USA
[2]California Institute of Technology, Pasadena, CA, USA
[3]Nvidia, Santa Clara, CA, USA

[*]Corresponding author: kong11@llnl.gov, Lawrence Livermore National Laboratory 7000 East Ave., Livermore, CA 94550-9234

**Abstract**

The recent development of Neural Operator (NeurOp) learning for solutions to the elastic wave equation shows promising results and provides the basis for fast large-scale simulations for different seismological applications. In this paper, we use the Fourier Neural Operator (FNO) model to directly solve the 3D Helmholtz wave equation for fast seismic ground motion simulations on different frequencies, and show the frequency bias of the FNO model, i.e. it learns the lower frequencies better comparing to the higher frequencies. To reduce the frequency bias, we adopt the multi-stage FNO training, i.e., after training a 1st stage FNO model for estimating the ground motion, we use a second FNO model as the 2nd stage to learn from the residual, which greatly reduced the errors on the higher frequencies. By adopting this multi-stage training, the FNO models show reduced biases on higher frequencies, which enhanced the overall results of the ground motion simulations. Thus the multi-stage training FNO improves the accuracy and realism of the ground motion simulations.

**Introduction**

Partial differential equations (PDE) play an important role in scientific and engineering simulations to help scientists understand complex physical systems or processes. Few of them can have analytical solutions, thus many numerical methods are developed to approximate the solutions, such as finite difference/element/volume methods, spectral element methods and so on (Evans et al., 2012; Simmons, 2016). Many of the numerical methods discretize both space and time demanding high-resolution grids to achieve accurate solution, therefore, even with the aid of high-performance computing (HPC), the heavy computational costs prohibit some applications, such as uncertainty quantifications, which relying on the generation of many simulations, real-time applications that require fast simulations, and so on.

Recent efforts on using the neural network-based approaches to speed up the simulations show promising results over the traditional numerical solvers (Blechschmidt & Ernst, 2021; Greenfeld et al., 2019; Kasim et al., 2021). More recently, a type of new data-driven approach called neural operators emerged as a great tool to solve PDEs. Comparing to the neural network-based approaches, the neural operator-based models can learn the mapping between infinite dimensional function spaces, thus achieve mesh-independence, i.e., trained model with the same parameters can be applied on different discretizations (Kovachki, Li, et al., 2021; Z. Li et al., 2020a, 2020b). In particular, the Fourier Neural Operator (FNO) and its variants demonstrated both computational efficiency as well as good accuracy in solving various PDEs comparing to the numerical methods based on HPC (Azizzadenesheli et al., 2024; Choy et al., 2025; Kovachki, Lanthaler, et al., 2021; Kovachki, Li, et al., 2021; Z. Li et al., 2020a; Tang et al., 2024). For example, (Wen et al., 2023) uses a nested FNO models for real-time CO2 geological storage prediction and achieves a speedup of 700,000 times comparing to existing numerical solvers. The FourCastNet (Kurth et al., 2023), which utilizes an adaptive FNO model, achieves a speedup 10,000 times in global weather forecasting. Furthermore, the mesh-independent property allows the trained FNO model to evaluate solutions on higher resolution grids than the training resolution grids.

Similar trend appears in the field of seismology. The seismic wave propagations in 3D Earth media are governed by the wave equations can be solved by numerical methods with heavy computational cost (Igel, 2017). Therefore, machine learning approaches are developed as surrogate models to speed up the simulations (Karimpouli & Tahmasebi, 2020; Moseley et al., 2020; Song et al., 2023; Song & Wang, 2022; Tamhidi et al., 2022). Recently, efforts to use the neural operator-based models show promising results in solving the wave equations (B. Li et al., 2023; Shi et al., 2024; Sun et al., 2022; Wei & Fu, 2022; Y. Yang et al., 2023; Y. Yang et al., 2021) both in the acoustic and elastic version, but due to the computational cost as well as the GPU memory usage limitation, most of these are on relatively small scales or 2D problems. There are also efforts to use the neural operators on 3D problems (Kong & Rodgers, 2023; Lehmann et al., 2024) in the time domain, as well as to decompose the seismic wavefield into the frequency domain using the Helmholtz equations to overcome the memory required for the 3D time domain equations (Zou et al., 2024).

Meanwhile, the machine learning communities identified many of the neural network models share the limitation of the frequency biases (Rahaman et al., 2019; Xu et al., 2024; Xu et al., 2019), which show that the neural networks are good at learning low-frequencies yet struggle in learning high-frequencies. Even the FNO models show similar frequency bias in solving the Navier-Stokes and Darcy equations (Qin et al., 2024). Different approaches are proposed to reduce this bias, and multi-stage training shows promising results on neural networks as well as FNO models (Qin et al., 2024; Wang & Lai, 2024). In this paper, we show that frequency biases also exist with FNO models in solving 3D seismic wave equations. Inspired by (Qin et al., 2024; Wang & Lai, 2024), we used multistage training, specifically similar to the so-called spectral boosted FNO (Qin et al., 2024) to reduce discrepancies in learning the high-frequencies. We show an improvement in the overall performance of FNO models, either to train the model with different frequencies, or one FNO model for each individual frequency, in the forward modeling. Meanwhile, we also show the potential improvement of the inversion of subsurface structure imaging as well.

**Methods**

*Elastic Wave Equations*

The propagation of the waves in the elastic media is governed by the second-order hyperbolic elastic wave equations. The 3D time domain wave equation can be transformed to the Fourier domain and solve as a set of Helmholtz equations:

$$\omega^2 \rho U_x + \frac{\partial}{\partial x}\left[\lambda\left(\frac{\partial U_x}{\partial x} + \frac{\partial U_y}{\partial y} + \frac{\partial U_z}{\partial z}\right) + 2\mu \frac{\partial U_x}{\partial x}\right] + \frac{\partial}{\partial y}\left[\mu\left(\frac{\partial U_x}{\partial y} + \frac{\partial U_y}{\partial x}\right)\right]$$

$$+ \frac{\partial}{\partial z}\left[\mu\left(\frac{\partial U_x}{\partial z} + \frac{\partial U_z}{\partial x}\right)\right] + F_x = 0$$

$$\omega^2 \rho U_y + \frac{\partial}{\partial y}\left[\lambda\left(\frac{\partial U_x}{\partial x} + \frac{\partial U_y}{\partial y} + \frac{\partial U_z}{\partial z}\right) + 2\mu \frac{\partial U_y}{\partial y}\right] + \frac{\partial}{\partial x}\left[\mu\left(\frac{\partial U_x}{\partial y} + \frac{\partial U_y}{\partial x}\right)\right]$$

$$+ \frac{\partial}{\partial z}\left[\mu\left(\frac{\partial U_y}{\partial z} + \frac{\partial U_z}{\partial y}\right)\right] + F_y = 0$$

$$\omega^2 \rho U_z + \frac{\partial}{\partial z}\left[\lambda\left(\frac{\partial U_x}{\partial x} + \frac{\partial U_y}{\partial y} + \frac{\partial U_z}{\partial z}\right) + 2\mu \frac{\partial U_z}{\partial z}\right] + \frac{\partial}{\partial x}\left[\mu\left(\frac{\partial U_x}{\partial z} + \frac{\partial U_z}{\partial x}\right)\right]$$

$$+ \frac{\partial}{\partial y}\left[\mu\left(\frac{\partial U_y}{\partial z} + \frac{\partial U_z}{\partial y}\right)\right] + F_z = 0$$

where $\rho$ is the material density, $\omega$ is the angular frequency, and $\widehat{U}_i$ and $\widehat{F}_i$ ($i = x, y, z$) are the displacements and body forces in the frequency domain. $\lambda$ and $\mu$ are the Lamé parameter and shear modulus related to the medium properties, which they can be linked to the P- and S-wave velocities through the relationship below.

$$\alpha = \sqrt{\frac{\lambda+2\mu}{\rho}}, \beta = \sqrt{\frac{\mu}{\rho}} \tag{2}$$

*Multistage FNO Structure and Training Setup*

We mainly used the original FNO model developed in (Z. Li et al., 2021), which is shown mathematically as a universal approximator for any arbitrary nonlinear continuous operators (Kovachki, Lanthaler, et al., 2021). The FNO model has the property of learning the mappings between infinite-dimensional function spaces, a key difference to the traditional neural network, which only maps finite dimensional vector spaces. Thus, in the context of solving a partial differential equation (PDE), the FNO $\mathcal{G}_\theta$ attempts to learn a mapping between the input field $\mathcal{A}$ that usually consists of a function space containing coefficients, initial/boundary conditions of the PDE and the coordinates as the features, and a function space $\mathcal{U}$ containing the output of the PDE

$$\mathcal{G}_\theta: \mathcal{A} \to \mathcal{U} \tag{3}$$

where $\theta$ is the parameterization of the FNO model that can be learned through the training. Figure 1 broadly sketches the experimental setup for this study.

As a neural network, the FNO model uses an iterative architecture to learn this mapping. The model first lifts the input function a(x), where it includes equation coefficients, initial/boundary conditions, forcing, grids etc., via a P operator into higher dimension space, which can be achieved by pointwise transformation on each point along the feature dimension. This lifted space subsequently passes through a series of FNO layers until a Q operator to project it back to the output dimension. The bottom part of Figure 1 shows a schematic of the FNO model structure. Each of the FNO layers can be written as

$$v_{k+1}(x) = \sigma(Wv_k(x) + (\mathcal{K}(\phi)v_k)(x)) \tag{4}$$

where $v_k$ is the k$^{th}$ layer, $\sigma$ is a nonlinear activation function, W is the linear transformation weights to be learned along the feature dimension, and $\mathcal{K}(\phi)$ is the kernel integral operator with the parameters $\phi$ to be learned. The so called FNO model employs Fourier Transform $\mathcal{F}$ to approximate the kernel operator, as shown in the following equation:

$$(\mathcal{K}(\phi)v_k)(x) = \mathcal{F}^{-1}(R_\phi \cdot (\mathcal{F}v_k))(x) \tag{5}$$

where $R_\phi$ is a restriction operator, which truncates the Fourier modes to lower frequencies, essentially a low-pass filter, with the pointwise weight parameter $\phi$ to be learned. A user can define how many Fourier modes to keep in each feature dimension. More detailed information about FNO can be found in the paper (Z. Li et al., 2021).

To solve the Helmholtz wave equation, we use discretized pairs of data (A, U), as shown in the upper part of figure 1, where A is the input data includes 3D P- and S-wave velocities, the amplitude distribution of the source in frequency domain, source characteristics (strike, dip and rake) as well as the frequency component (omega in the figure) we are solving for. The solution from the PDE is U, which consists of 3-component real and imaginary parts of the displacements at each of the spatial grid (64x64x64) that generated using the spectral element method (see more details in the Data section). The normal flow of the forward estimation using the FNO model is shown in Figure 1. The estimation of the 3-component ground motions on the 3D volumes by the FNO model (with the parameters initialized randomly) is used to obtain the errors matching the ground truth Y. The loss function we used here is a combination of pointwise L2 and L1 norm, which is the same as that used in (Y. Yang et al., 2023; Y. Yang et al., 2021; Zou et al., 2024), as shown below:

$$loss = 0.9L_1 + 0.1L_2 \qquad (6)$$

This loss can be propagated back through the network via an Adam optimizer (Kingma and Ba, 2017), we use 0.01 as the learning rate, and shrink it to half every 30 epochs to update the FNO model. As for the FNO model, we used the original FNO structure described in (Z. Li et al., 2021), with width 48 and modes 30 for the X, Y, Z direction.

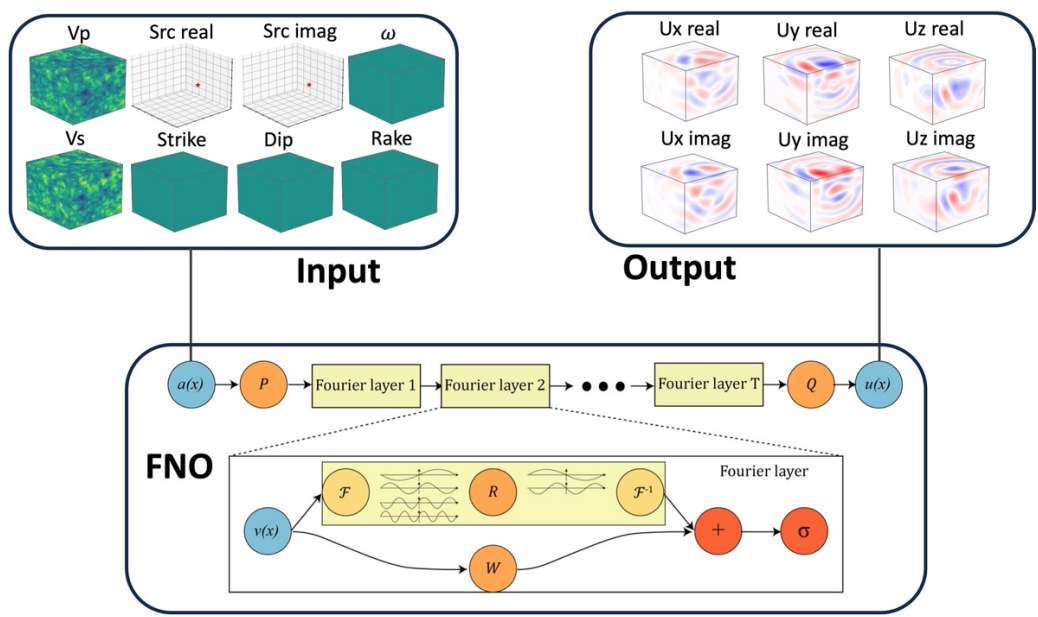

**Figure 1.** The FNO model for seismic forward simulation. The 3D P- and S-wave speeds (Vp and Vs), the source distribution in frequency domain, strike, dip, rake, as well as the angular frequency (ω) are used as the input which are fed into the FNO model to approximate the 3D ground motions in frequency domain as output.

*Multistage FNO model*

Figure 2 shows the setup of a 2-stage FNO model to improve the model's capability on modeling the high-frequency contents, reducing the frequency biases that present using the FNO model. Stage 1 trains a FNO model as we described in the previous section, the only difference is that after getting the final trained model and the estimated ground motion $\widehat{U}$, we compute the residuals by comparing it to the ground truth $U$. The frequency dependent residuals from stage 1 are used as the target to be fit by the stage 2 model (see figure 3). Thus, in stage 2 we have another FNO model to try to capture the residuals left from stage 1. The input for the stage 2 model includes both the input and output from the stage 1 model, which provide information about the output status of stage 1. With the guidance of the residuals from stage 1, $R$, the FNO model in the stage 2 estimates the residuals $\widehat{R}$ that missed by the previous model, mostly from high frequencies and a little from low frequencies as well. After the training of stage 2 model, the final solution for the wave equation, can be represented by the sum of the results from both of the stages, i.e.

$\widehat{U} + \widehat{R}$. We use the same FNO structure, loss function as well as the optimization algorithm through the two stages.

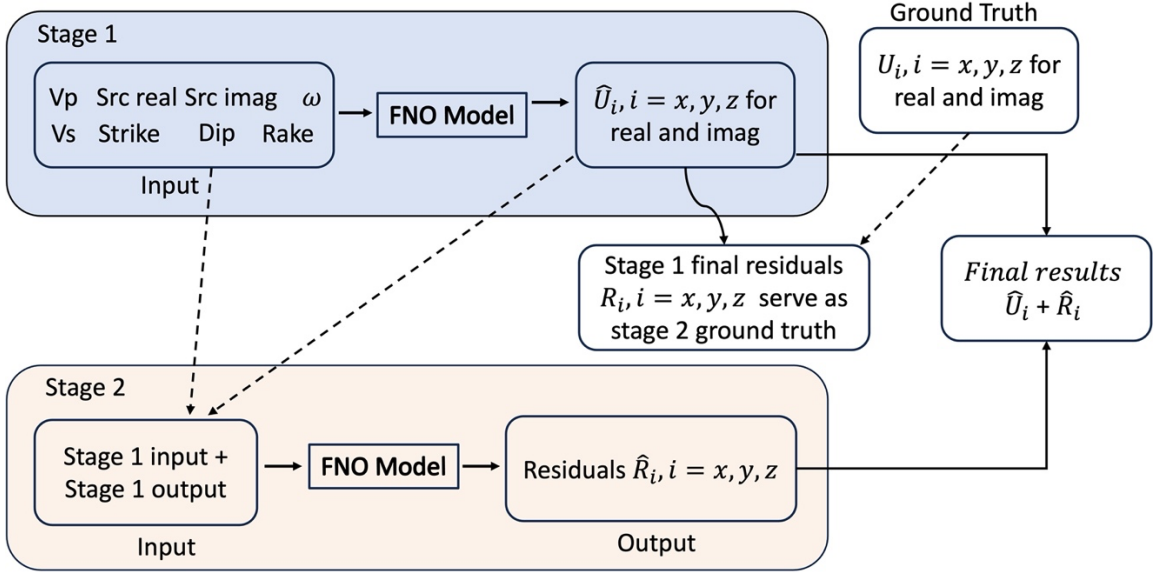

**Figure 2.** Setup of the multistage FNO model. The blue shaded box contains stage 1 while the orange shaded box contains the stage 2 input and output of the FNO model. The final results are a combination of the output from the two stages.

*Data Parallel on Multi-nodes*

The overall pipeline is implemented in a data parallel fashion to use a multi-node computing environment. We use LLNL's Tioga computing platform, each node consists 8 AMD MI-250X GPU (128 Gb memory each) for training the FNOs. For training the a FNO model in each stage for a single frequency (the FNO-11 model described in the discussion session), we use 5 nodes with 40 GPUs in total, while for training the FNO model for all frequencies (the FNO-1 model described in the discussion session), we use 12 nodes with 96 GPUs to accelerate the training.

**Data**

A 3D seismic wave propagation simulation dataset is generated using Salvus, a spectral element method (Afanasiev et al., 2019), for model training with a split 15,000, 1000, 500 as training, validation, and testing. Each data simulation is on a discretized 3D domain with a 64x64x64 grids and 5 km in X, Y, Z direction. As suggested by the Salvus documentation, we take one element per wavelength at the maximum frequency and a 4$^{th}$ order polynomial for the simulations. This computational domain contains a free-surface boundary condition at the top boundary with other five sides as absorbing boundary conditions. For each simulation the P- and S-wave velocities, and a randomly located earthquake source in the domain with strike, dip and rake randomly draw from 0-360, 0-90, 0-180 degrees are specified. We use a Ricker wavelet with a central frequency of 3 Hz as the source time function. We use a background of 3 km/s with stochastic Von Kármán random field perturbations (Hurst exponent of 0.5, correlation length of 8 grid cells) with a standard deviation of 10% to generate the velocities of S-wave (Vs) fields. The velocity ratio between P and S (Vp/Vs) are generated by perturbing a background ratio of 1.732 with the Gaussian random fields, which have 32 grid cells as the correlation length and a standard deviation of 2%. Thus, the P-wave velocity can be computed as Vs multiply the Vp/Vs ratio. A constant density 2.7 g/cm$^3$ is utilized in the simulations. We use 0.002 s time step for the simulations to ensure the Courant-Friedrichs-Lewy condition (Courant et al., 1967). The 3-component displacements are then sampled to the 64x64x64 grid points in 3D for a total of 2.5s with a time step 0.04s. After obtaining the Salvus simulations, the source time function as well as the time series of the displacement fields are transformed into the Fourier domain and 11 discrete frequencies (2 – 6 Hz with 0.4 Hz step) are used for the Helmholtz equation, which we eliminate the frequencies with little energy in the wavefield spectrum. The P- and S-wave velocity distributions, the complex source term indicating source location, source characteristics (strike, dip and rake), and the frequency desired are fed into the FNO model to train it to generate the 3D wavefield (complex values) as shown in Figure 1.

**Results**

We start our experiment by training individual FNO models (as shown in figure 1), each focused on the solution within a separate frequency band. When the results are summed together from the different models, these solutions should reproduce the full wavefield. Figure 3 shows the frequency bias from the FNO model, where the model tends to learn how to reproduce the lower frequencies more quickly than the higher frequencies. Part of the reasons attribute to the physics, that higher frequency contents correspond to wavefields with shorter wavelengths, thus harder to model. The other part of the reasons attributed to the tendency of neural network types of the model tend to generate smooth output using the non-linear activation function to increase the generalization capability of the model, thus causing the model to fit the lower frequency content easier (Neal et al., 2018; Xu et al., 2024; Z. Yang et al., 2020).

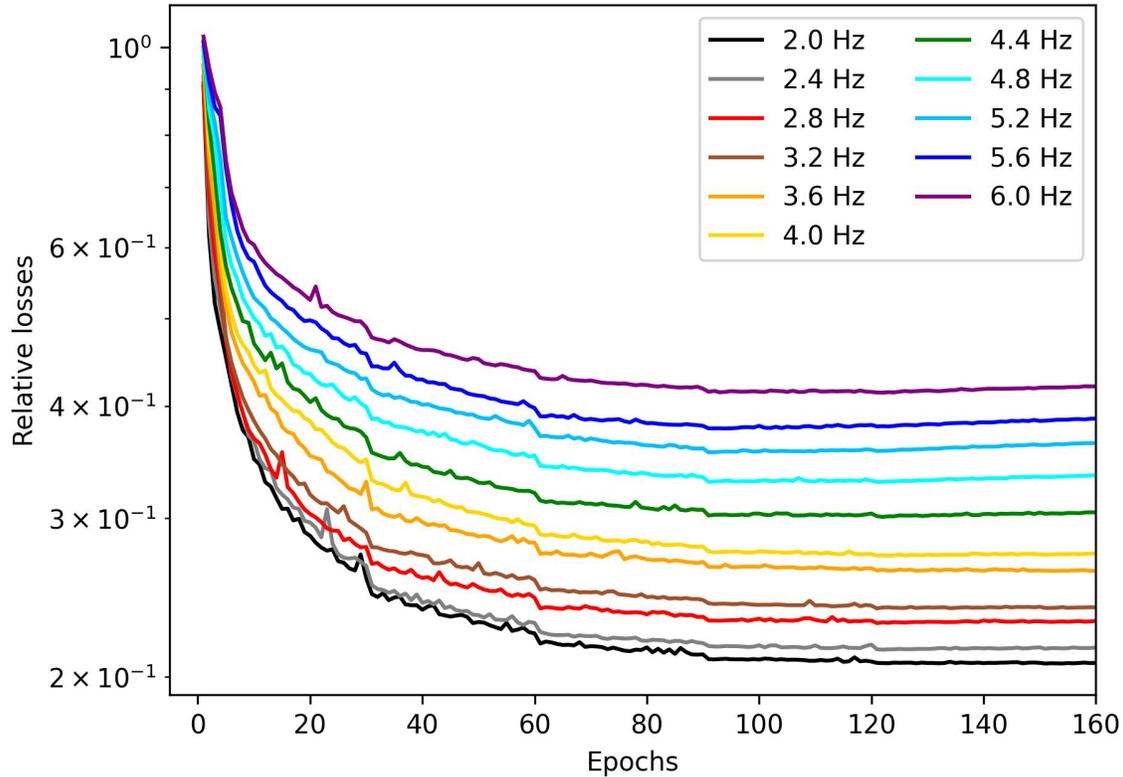

**Figure 3.** Relative validation losses versus training epoch for different frequencies training. Each FNO model is trained using single frequency data (15,000 samples) and validated on 1,000 samples.

To address the frequency bias of the FNO model in seismic forward simulations, we adopt the multi-stage training (Qin et al., 2024; Wang & Lai, 2024), i.e. after the FNO models finishes the estimation of the ground motion, the residuals between the estimation and target are calculated, which become the target of the 2nd stage of training new FNO models to model these residuals. The idea is that the 1st stage FNO models focus on fitting the main characteristics of the data, the small residuals for the 2nd stage will be more highlighted in the training data thus easier to be fitted for a new FNO model. To illustrate the effectiveness of the 2nd stage training, figure 4 shows the validation curves for 3 frequencies training for both stages. We only show 3 frequencies here, but the other frequencies do have similar trends. We can see after the convergance of the stage 1 training, stage 2 models can further reduce the errors that left in the previous training. Furthermore, the higher the frequencies, the more errors can be reduced by the stage 2 models. This indicates that the lower frequencies are relatively easy to fit by the stage 1 model, thus not leave too much room to improve on as opposed to the higher frequencies. Figures 5 and 6 show some specific examples from test data (source at 22, 50, 25) between the performance of single stage training versus the multi-stage training. In both cases, we can see the residuals from the multi-stage training are smaller comparing to the single-stage training, the higher frequency changes have better fitting.

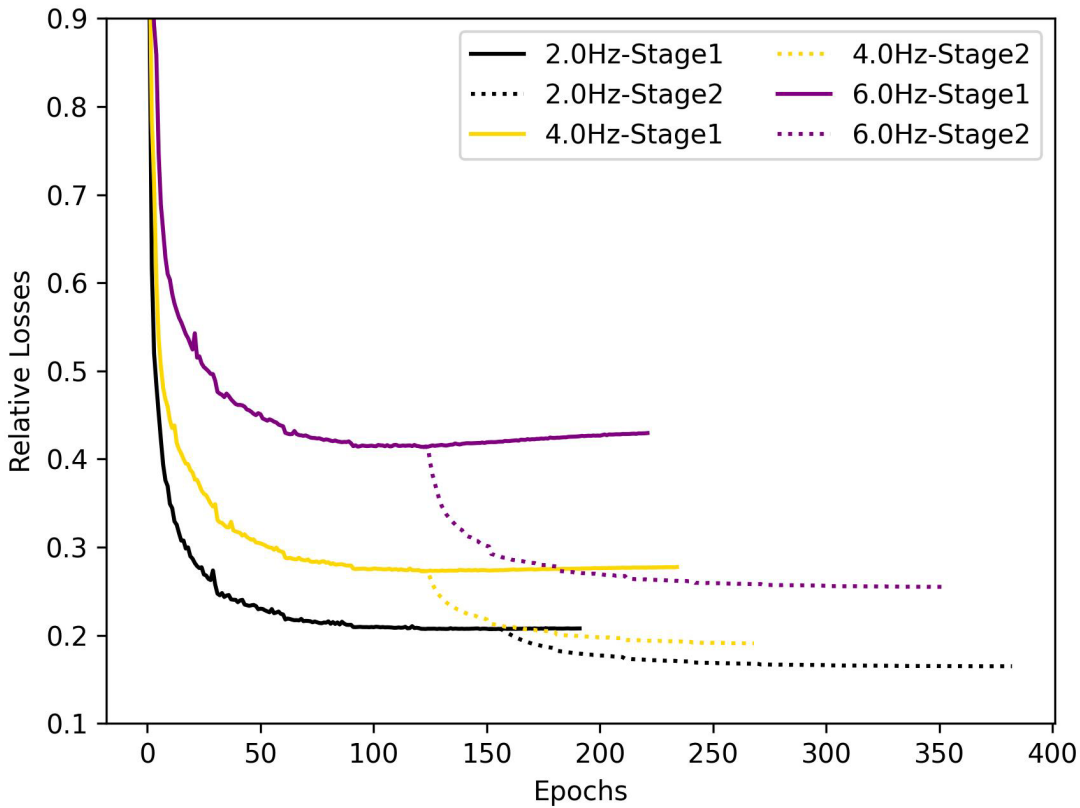

**Figure 4.** Multi-stage training validation curves with relative losses. The solid lines are the stage 1 training while the dotted lines are from the stage 2 training colored by the frequencies. The stage 2 trainings start from the best trained models from stage 1.

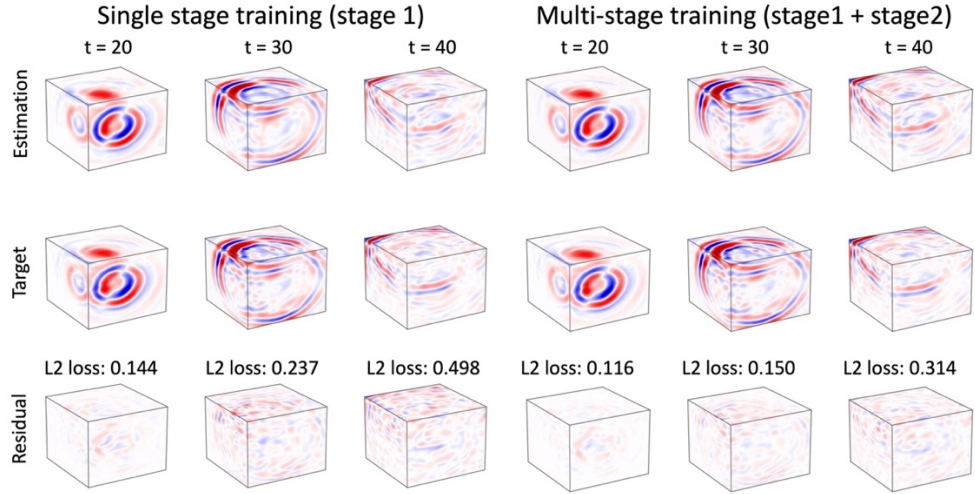

**Figure 5.** Ground motion snapshots at t = 10, 20, 30. The left 3 columns show the stage 1 results while the right 3 columns show the multi-stage training results. The estimation from FNO models, target and residuals are shown in the 3 rows. Relative L2 loss is shown at the top of the last row for each snapshot.

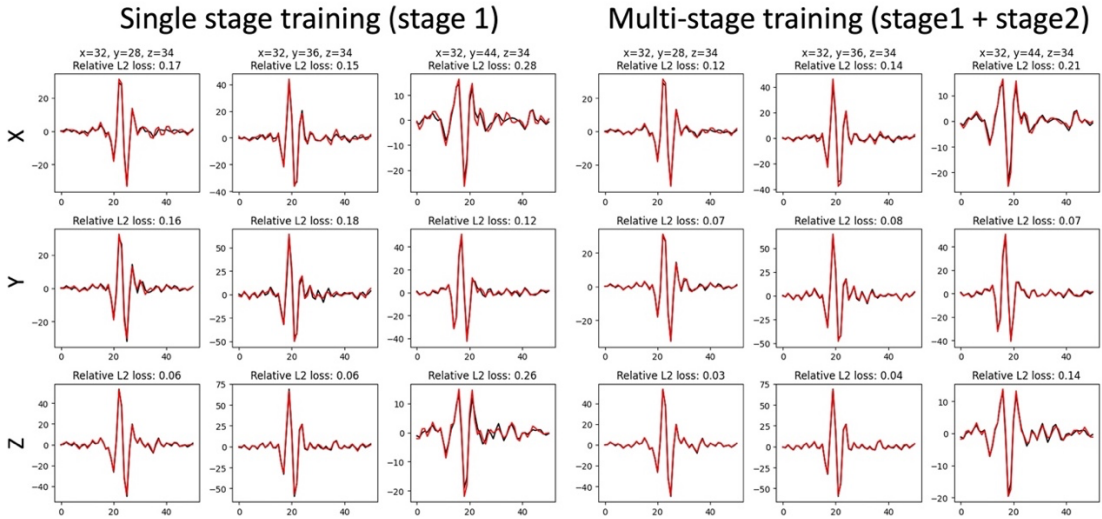

**Figure 6.** The comparison of the 3-component seismic waveforms at different grid locations. The Red lines and black lines are results from the FNO model and that of the spectral element solver respectively. The top row titles show the grid locations (x, y, z). The relative L2 losses between the FNO estimation and the spectral element solver are shown in the title as well. In each panel, horizontal axis is the time,

while vertical axis is the amplitude. X, Y, and Z components are shown in the 3 rows. The left 3 columns show the stage 1 results while the right 3 columns show the multi-stage training results.

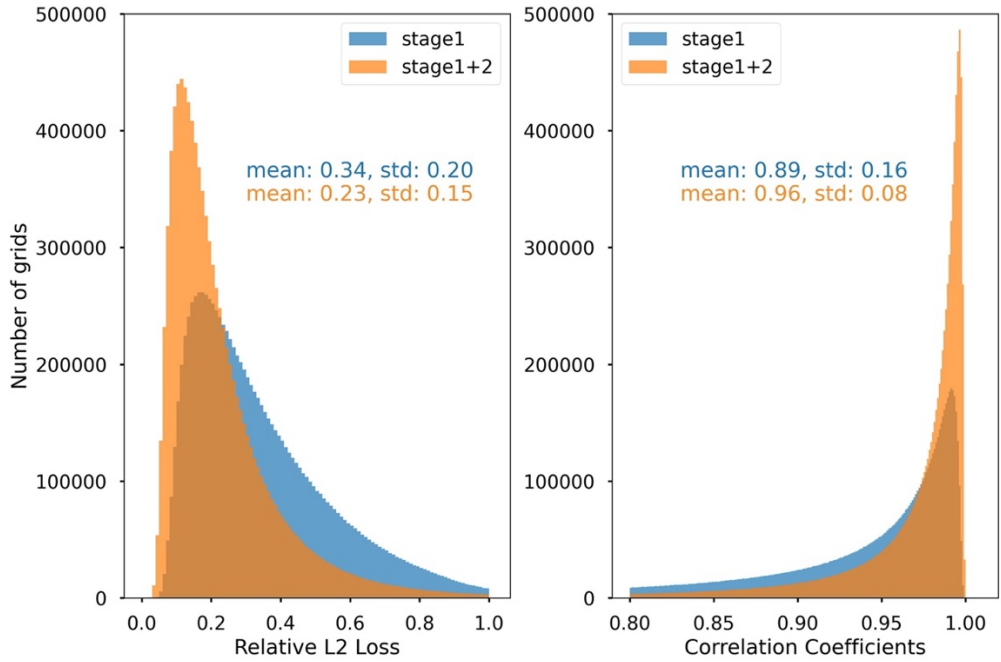

**Figure 7.** Distribution of Relative L2 loss as well as the correlation coefficient for every other grid in the domain for all the 3 components. Stage 1 model results are shown in blue color while the multi-stage model results are shown in orange color, with mean and standard deviation are shown in the texts. The bins for relative L2 losses are 0 to 1 with 0.01 as the bin width, while 0.8 to 1 with 0.001 for correlation coefficients.

We use 100 test simulations to evaluate the performance of the trained model on every other grid. On each of these grids, the metrics of the time series of the x, y, z components are computed and then aggregated to generate the histograms as shown in Figure 7, which shows the relative L2 loss and correlation coefficients. Stage 1 metrics are shown in blue, while stage 1+2 metrics are shown in orange. The multi-stage training improves both metrics, i.e. reducing the mean of the L2 loss by about 1/3 and improving the correlation coefficients from 0.89 to 0.96. To show the metrics on each frequency, we flatten the corresponding frequency output and simulation data into vectors, and compute the relative L2 loss and correlation coefficients directly, and plot the mean and standard deviation against frequencies in Figure 8. We can clearly see the trend if we only use stage 1 training, for example, the relative L2 losses are getting higher with the increase of the frequencies, while the correlation coefficients are generally dropping. The multi-stage training shows a similar trend, but this trend is very mild, and the relatively L2 losses only slightly increasing with frequency. This all demonstrate the effectiveness of the multi-stage training.

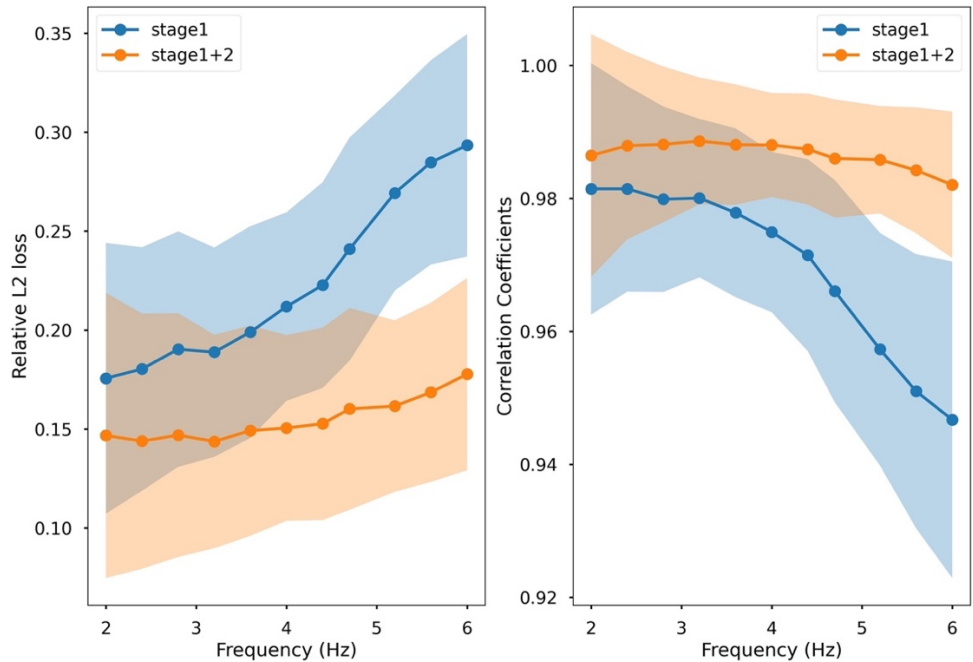

**Figure 8.** Performance metrics against frequencies for single-stage and multi-stage training. Left panel shows the relative L2 losses while the right panel shows the correlation coefficients. Dots with solid lines show the mean values for each frequency, while the shaded areas are the standard deviations.

**Conclusion and Discussion**

In this paper, we use the FNO model to solve the 3D elastic wave equation in the frequency domain with earthquake source represented by using the strike, dip and rake. Using 15,000 simulations from spectral element methods, the FNO models are trained in a supervised way to estimate the ground motion. The trained FNO models show the tendency of fitting the lower frequency better but not the higher frequencies. By using the multi-stage training, where new FNO models are used to fit the residuals from the first stage FNO models, the errors in the higher frequencies can be reduced and the overall performance can be improved. The test results show with the multi-stage training, the errors across different frequencies are mostly flat, with only a slight increase with frequencies.

To further test the multi-stage training, we also test a stage 3 by using new FNO model to train on the stage 2 residuals in the hope that it will reduce the error even more. Figure 9a shows the stage 3 validation error for 6.0 Hz test, we can see it drops only about 4% which is not significant, thus we decide only to use 2-stage training.

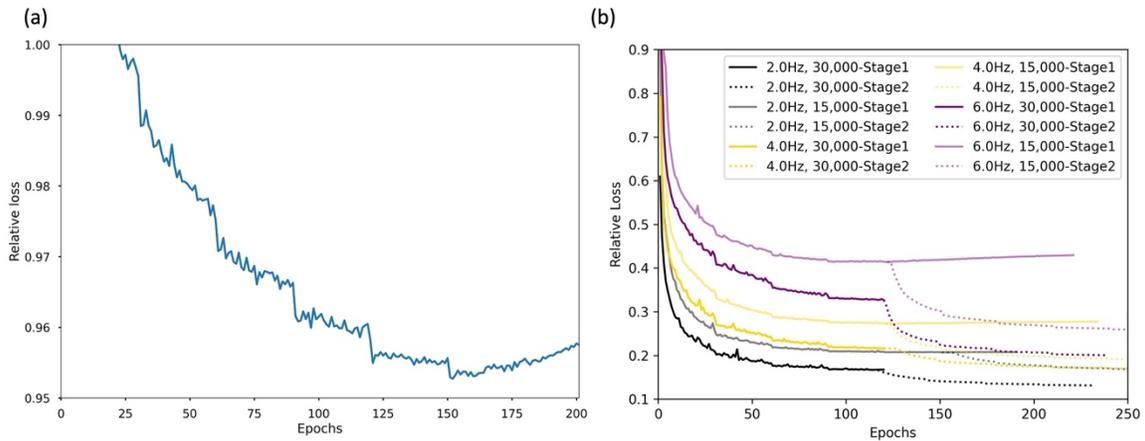

**Figure 9.** (a) Validation loss with training epochs for 6.0 Hz FNO model for stage 3. (b) Comparison of multi-stage training using 15,000 and 30,000 simulations for frequencies 2.0, 4.0 and 6.0 Hz. The solid lines are validation curves for stage 1 while dotted lines are for stage 2.

The main purpose of the paper is to show the frequency biases of FNO models can be reduced by using the multi-stage training, thus we only use 15,000 samples to train the FNO models for the consideration of computational cost. We could use more training data to reduce the error even more if we need a better model, Figure 9b shows the 2-stage training validation losses for 3 different frequencies using 30,000 simulations. It shows we can further reduce the error by adding more data.

In each stage, we use 11 FNO models to learn each of the 11 frequencies (let's call these models as FNO-11 models), in contrast of using only one FNO model for all the frequencies (let's call this model FNO-1). One consideration we have is that individual model that specifically learning only one frequency may capture better characteristics of the wavefield. The second consideration is mainly for computational purposes, it is easier to train 11 models since each model only sees 1/11 of the data, the training can be done with relatively small number of GPUs, which is easier to request in a multi-user cluster environment. For example, to train the FNO-11, the training takes about 200s per epoch with 40 GPUs, while to train the FNO-1 model for all the 11 frequencies, the training takes about 800s per epoch with 96 GPUs. Thus requiring smaller sets of GPUs with shorter running time is a big advantage on multi-user cluster without waiting in a queue too long. In order to test if there are performance differences between training FNO-11 and FNO-1, we further train one FNO-1 model to learn all the frequencies with 15,000 simulations. Figure 10 shows the test relative L2 losses on every other grid for 100 simulations for the trained FNO-1 model. Comparing to Figure 7 and 8, we can see the relative L2 losses histograms are basically the same, while the relatively L2 losses on different frequencies is a little higher for the FNO-1 model at high frequencies end. Thus this illustrates the FNO-11 and FNO-1 models are very similar, but the FNO-11 models perform slightly better, this may be due to simpler tasks for the model, i.e. only paying attention to one frequency at each time.

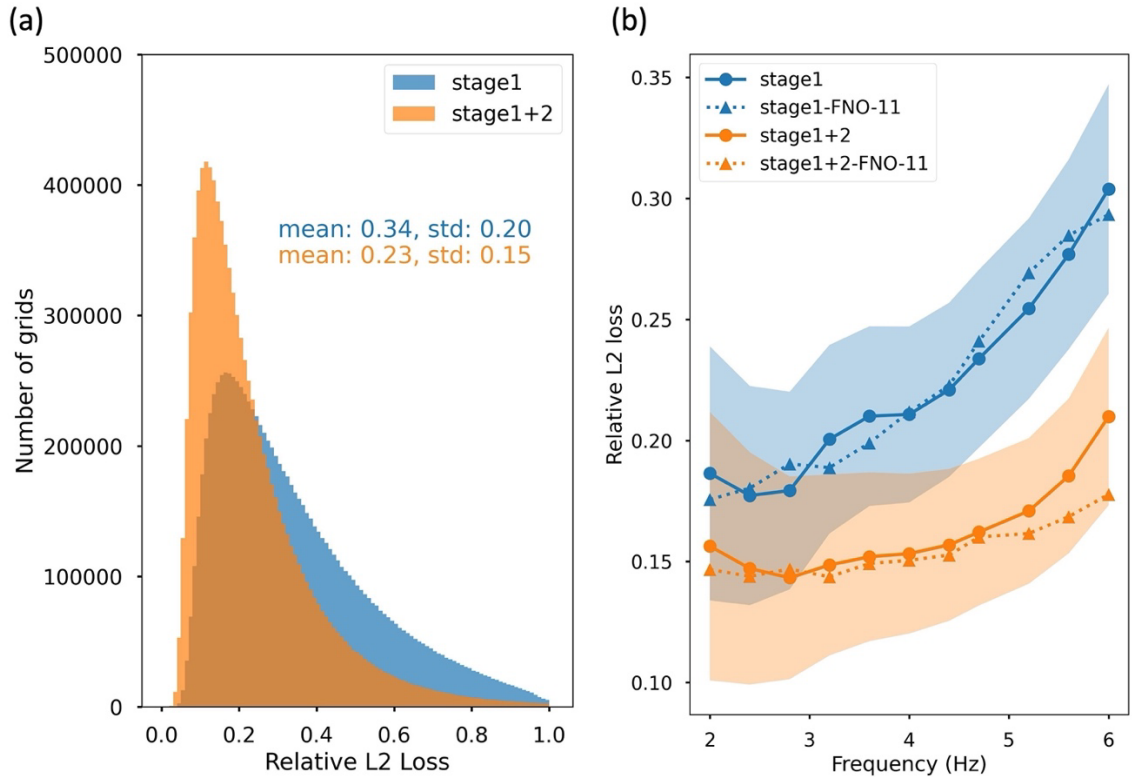

**Figure 10.** Performance metrics for only training one FNO model (FNO-1) for all frequencies. (a) Histograms of relative L2 losses for FNO-1 model. The bins are from 0 to 1 with 0.01 as bin width. The mean and standard deviation for the FNO-11 model are the same as shown here (b) Relative L2 losses against frequencies for FNO-1 model. Dots with solid lines show the mean values for each frequency, while the shaded areas are the standard deviations. The triangles and dotted lines are from mean values of the FNO-11 models for comparison.

**Data and Resources**

The datasets used to train the FNO model are generated by using the Salvus. The main code developed for this research can be obtained at https://github.com/qingkaikong/XXX (add the link after paper is final).

**Acknowledgements**

This research has been funded by the Laboratory Directed Research & Development at Lawrence Livermore National Laboratory, under 24-ERD-012. This research was performed in part under the auspices of the U.S. Department of Energy by the LLNL under Contract Number DE-AC52-07NA27344. This is LLNL Contribution LLNL-JRNL-2002877. We also thank the useful discussions by colleagues from LLNL, Tim Lin, Claire Doody, Bill Walter, Stephen C. Myers, Arben Pitarka, Hewei Tang, Joshua A. White, Graham Bench.

**Competing interests**

None.

Yang, Y., Gao, A. F., Azizzadenesheli, K., Clayton, R. W., & Ross, Z. E. (2023). Rapid Seismic Waveform Modeling and Inversion With Neural Operators. *IEEE Transactions on Geoscience and Remote Sensing, 61*, 1-12. https://ieeexplore.ieee.org/abstract/document/10091544

Yang, Y., Gao, A. F., Castellanos, J. C., Ross, Z. E., Azizzadenesheli, K., & Clayton, R. W. (2021). Seismic Wave Propagation and Inversion with Neural Operators. *The Seismic Record, 1*(3), 126-134. https://doi.org/10.1785/0320210026

Yang, Z., Yu, Y., You, C., Steinhardt, J., & Ma, Y. (2020). *Rethinking bias-variance trade-off for generalization of neural networks.* Paper presented at the International Conference on Machine Learning.

Zou, C., Azizzadenesheli, K., Ross, Z. E., & Clayton, R. W. (2024). Deep neural Helmholtz operators for 3-D elastic wave propagation and inversion. *Geophysical Journal International, 239*(3), 1469-1484. https://doi.org/10.1093/gji/ggae342

24